\documentclass{appolb}
\usepackage{epsfig}

\begin{document}
\title{From FAIR to RHIC, hyper clusters and an effective strange EoS for QCD%
\thanks{Presented at the international conference on Strangeness in Quark Matter 2011}%
}
\author{J.~Steinheimer$^{1}$, A.~Botvina$^{1,4}$, K.~Gudima$^{1,5}$, I.~Mishustin$^{1,6}$, S.~Schramm$^{1}$, M.~Bleicher$^{1}$, H.~St\"ocker$^{1,2,3}$
\address{$^{1}$ FIAS, Johann Wolfgang Goethe University,
Frankfurt am Main, Germany}
\address{$^{2}$ Institut f\"ur Theoretische Physik, Goethe-Universit\"at, Max-von-Laue-Str.~1,
D-60438 Frankfurt am Main, Germany}
\address{$^{3}$ GSI Helmholtzzentrum f\"ur Schwerionenforschung GmbH, Planckstr.~1, D-64291 Darmstadt, Germany}
\address{$^{4}$Institute for Nuclear 
Research, Russian Academy of Sciences, 117312 Moscow, Russia}
\address{$^{5}$Institute of Applied Physics, Academy of Sciences of Moldova, 
MD-2028 Kishinev, Moldova} 
\address{$^{6}$Kurchatov Institute, Russian Research Center,
123182 Moscow, Russia}
}
\maketitle
\begin{abstract}
Two major aspects of strange particle physics at the upcoming FAIR and NICA facilities and the RHIC low energy scan will be discussed.
A new distinct production mechanism for hypernuclei will be presented, namely the production abundances for hypernuclei from $\Lambda$'s absorbed in the spectator matter in peripheral heavy ion collisions.
As strangeness is not uniformly distributed in the fireball of a heavy ion collision, the properties of the equation of state therefore depend on the local strangeness fraction. The same, inside neutron stars strangeness is not conserved and lattice studies on the properties of finite density QCD usually rely on an expansion of thermodynamic quantities at zero strange chemical potential, hence at non-zero strange-densities. We will therefore discuss recent investigations on the EoS of strange-QCD and present results from an effective EoS of QCD that includes the correct asymptotic degrees of freedom and a deconfinement and chiral phase transition.
\end{abstract}
  
\section{Introduction}
The objective of the low energy heavy ion collider programs, at the RHIC facility on Long Island and the planned projects NICA in Dubna and FAIR near the GSI facility, is to find evidence for the onset of a deconfined phase \cite{Gyulassy:2004zy,Hohne:2005qm}. At the highest RHIC energies, experiments \cite{Adams:2005dq,Back:2004je,Arsene:2004fa,Adcox:2004mh} have already confirmed a collective behavior of the created system, signaling a change in the fundamental degrees of freedom. Lattice QCD calculations indeed expect a deconfinement crossover to occur in systems created at the RHIC. As theoretical predictions on the thermodynamics of finite density QCD are difficult (see e.g. \cite{Fodor:2002km,Allton:2002zi,Laermann:2003cv,deForcrand:2008vr}), one hopes to experimentally confirm a possible first order phase transition and consequently the existence of a critical endpoint, by mapping out the phase diagram of QCD in small steps.
Hadronic bulk observables which are usually connected to the onset of deconfinement are the particle flow and its anisotropies as well as particle yields and ratios \cite{Ollitrault:1992bk,Rischke:1996nq,Sorge:1996pc,Heiselberg:1998es,Scherer:1999qq,Soff:1999yg,Brachmann:1999xt,Csernai:1999nf,Zhang:1999rs,Kolb:2000sd,Bleicher:2000sx,Stoecker:2004qu,Zhu:2005qa,Petersen:2006vm,Gazdzicki:2004ef,Gazdzicki:1998vd}. It has often been proposed, that e.g. the equilibration of strangeness would be an indication for the onset of a deconfined phase, although this idea is still under heavy debate \cite{Koch:1986ud,249179,Bratkovskaya:2000eu,BraunMunzinger:2003zz,Greiner:2004vm}.
Two main aspects of strangeness physics, closely connected to the equilibration of strangeness and the hyperon interactions, are the formation of nuclear clusters with strange content and the bulk properties of very dense nuclear matter with finite strangeness content.

\section{Hypernuclei}
Exotic forms of deeply bound objects with strangeness have been proposed \cite{Bodmer:1971we} 
as states of matter, either consisting of baryons or quarks.  
The H di-baryon was predicted by Jaffe
\cite{Jaffe:1976yi} and later, many more bound di-baryon states with strangeness were proposed
using quark potentials \cite{Goldman:1987ma,Goldman:1998jd} or the Skyrme model \cite{Schwesinger:1994vd}.
However, the non-observation of multi-quark bags, e.g. strangelets is still one of the open problems of 
intermediate and  high energy physics.
On the hadronic side, hypernuclei are known to exist and be produced in heavy Ion collisions already for a long time \cite{nucl-th/9412035,Ahn:2001sx,Takahashi:2001nm,arXiv:1010.2995}. 
Metastable exotic multi-hypernuclear objects (MEMOs)
as well as purely hyperonic systems of $\Lambda$'s and $\Xi$'s
were introduced in \cite{Schaffner:1992sn,Schaffner:1993nn} as the hadronic counterparts to
multi-strange quark bags \cite{Gilson:1993zs,SchaffnerBielich:1996eh}.
A motivation of hypernuclear physics is that it offers a direct
experimental way to study hyperon--nucleon ($YN$) and hyperon--hyperon
($YY$) interactions ($Y=\Lambda,\Sigma,\Xi,\Omega$). 
The nucleus serves as a laboratory offering the unique opportunity 
to study basic properties of hyperons and their interactions.

\subsection{Hypernuclei production in the spectator fragments}

\begin{figure}[t]
 \centering
 \includegraphics[width=0.5\textwidth]{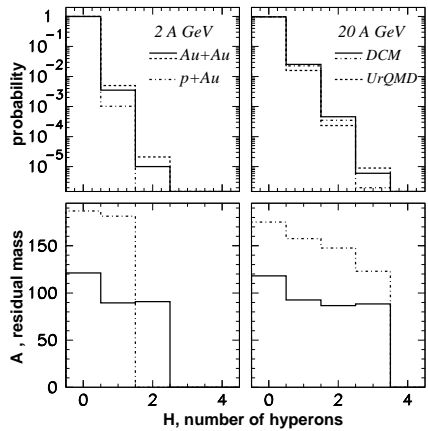}
 \caption{Probability per event for the formation of conventional and strange spectator residuals 
(top panels), and their mean mass numbers (bottom panels)
versus the number of captured $\Lambda$ hyperons (H), calculated with DCM 
and UrQMD model for p + Au and  Au + Au collisions with energy of 
2 GeV per nucleon (left panels), and 20 GeV per nucleon (right panels). 
The reactions and energies are noted in the figure by different 
histograms.}
 \label{rates}
\end{figure}

In this work we will focus on the production of hypernuclei in high energy collisions of Au+Au ions. In such systems strangeness is produced abundantly and is likely to form clusters of different sizes. We can discriminate two distinct mechanisms for hypercluster formation in heavy ion collisions. First, the formation of hypernuclei in the hot and dense fireball of most central heavy ion collisions where the general assumption is that hypernuclei are formed at or shortly after the hadronisation/chemical freeze out of the hadrons produced. In this work we will focus on a different production mechanism, the absorption of hyperons in the spectator fragments of non central heavy ion collisions. In this scenario we are interested in 
hyperons which propagate with velocities close to the initial velocities of the nuclei, i.e., in the vicinity of nuclear spectators. To calculate the absorption rate we employed the Ultra-relativistic Quantum Molecular Dynamics model (UrQMD v2.3)~\cite{Bleicher:1999xi,Bass:1998ca} and the intra-nuclear cascade model (DCM) developed in Dubna 
\cite{195629} to estimate the model dependence of the predictions. The hyperons produced in the hot and dense stage of a heavy ion collisions can be absorbed by the spectators if their 
kinetic energy in the rest frame of the residual nucleus is lower than the 
attractive potential energy, i.e., the hyperon potential given by \cite{Ahmad:1983re}: 
\begin{equation}
	V_{\Lambda}(\rho)= -\alpha \frac{\rho}{\rho_0} 
     \left[1-\beta (\frac{\rho}{\rho_0})^{2/3}\right] ,
\label{binlam}
\end{equation}
where $\alpha=57.5$ MeV, and $\beta =0.522$.  
The local nucleon density $\rho$ at the hyperon's position is calculated within the hadronic transport models, whereas the details of the computation and more results on the properties of the absorbed hyperons can be found in \cite{arXiv:1105.1341}.
Figure \ref{rates} shows the resulting probabilities for the formation of a conventional and strange spectator residual 
(top panels), and their mean mass numbers (bottom panels)
versus the number of captured $\Lambda$ hyperons (H), calculated with the DCM 
and UrQMD models for p + Au and  Au + Au collisions at an energy of 
2 GeV per nucleon (left panels), and 20 GeV per nucleon (right panels). One clearly observes that the production of heavy multi-hyper nuclei is possible at FAIR.
 
\section{The strange equation of state}
The strange EoS is of particular interest for the understanding of several aspects of QCD:
\begin{enumerate}
\item As has been shown in \cite{Steinheimer:2008hr} the net strangeness distribution in phase space of a heavy ion collision can fluctuate, although the total net strangeness is zero. To dynamically treat such a system, the equation of state for $\rho_{s}\ne 0$ needs to be evaluated.
\item Compact stars are very dense and long lived objects. Due to a $\beta$-equilibrium inside the star, net-strange conservation is violated by the weak interaction.
\item Lattice QCD results at finite $\mu_B$ are often evaluated through a Taylor expansion in $\mu_B$ at $\mu_B=\mu_S=0$. A vanishing strange number chemical potential induces a non-vanishing net strangeness, which means that the equation of state of net-strange matter is calculated.
\end{enumerate}

First investigations on the strange equation of state were done in \cite{Lee:1992hn}, where one usually considered a first order transition from a hadron to a quark phase. In our study we employ the recently developed SU(3)$_f$ parity doublet for hadronic matter and its extension to quark degrees of freedom. In this approach an explicit mass term for baryons is possible, where
the signature for chiral symmetry restoration is the degeneracy of the baryons and their respective
parity partners. Adding an effective quark and gluon contribution is done via a PNJL-like approach \cite{Fukushima:2003fw,Ratti:2005jh}. This model uses the Polyakov loop $\Phi$ as the order parameter for deconfinement. $\Phi$ is defined via $\Phi=\frac13$Tr$[\exp{(i\int d\tau A_4)}]$, where $A_4=iA_0$ is the temporal component of the SU(3) gauge field. To suppress the hadronic contributions from the equation of state at high temperatures and densities, effects of finite-volume hadrons are included in a thermodynamic consistent way. This model allows for a smooth transition from a hadronic to a quark dominated system, where the order parameters and thermodynamic quantities are in reasonable agreement with recent lattice data. For a detailed description of the parity model and comparisons with lattice we refer to \cite{arXiv:1108.2596}.

\begin{figure*}[t]
 \centering
\includegraphics[width=\textwidth]{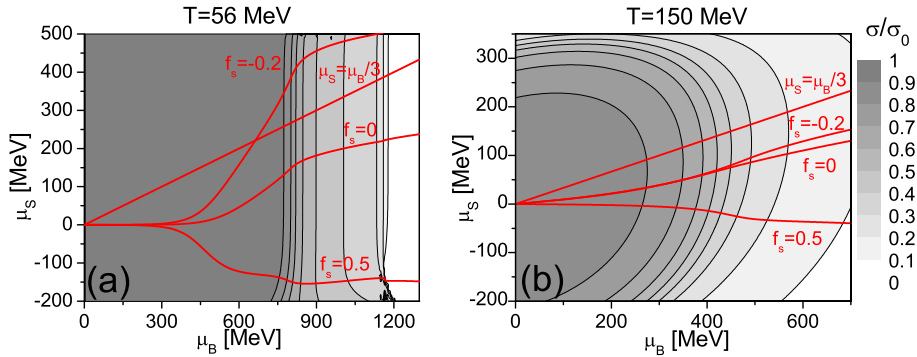}
 \caption{(Color online) Contour plots of the normalized chiral condensates as a function of the chemical potentials $\mu_B$ and $\mu_S$ for fixed temperature (a: $T=56$ MeV, b: $T= 150$ MeV). The red lines correspond to different values of a fixed strangeness to baryon fraction $f_s$.}
 \label{11}
\end{figure*}
\begin{figure}[t]
 \centering
\includegraphics[width=0.6\textwidth]{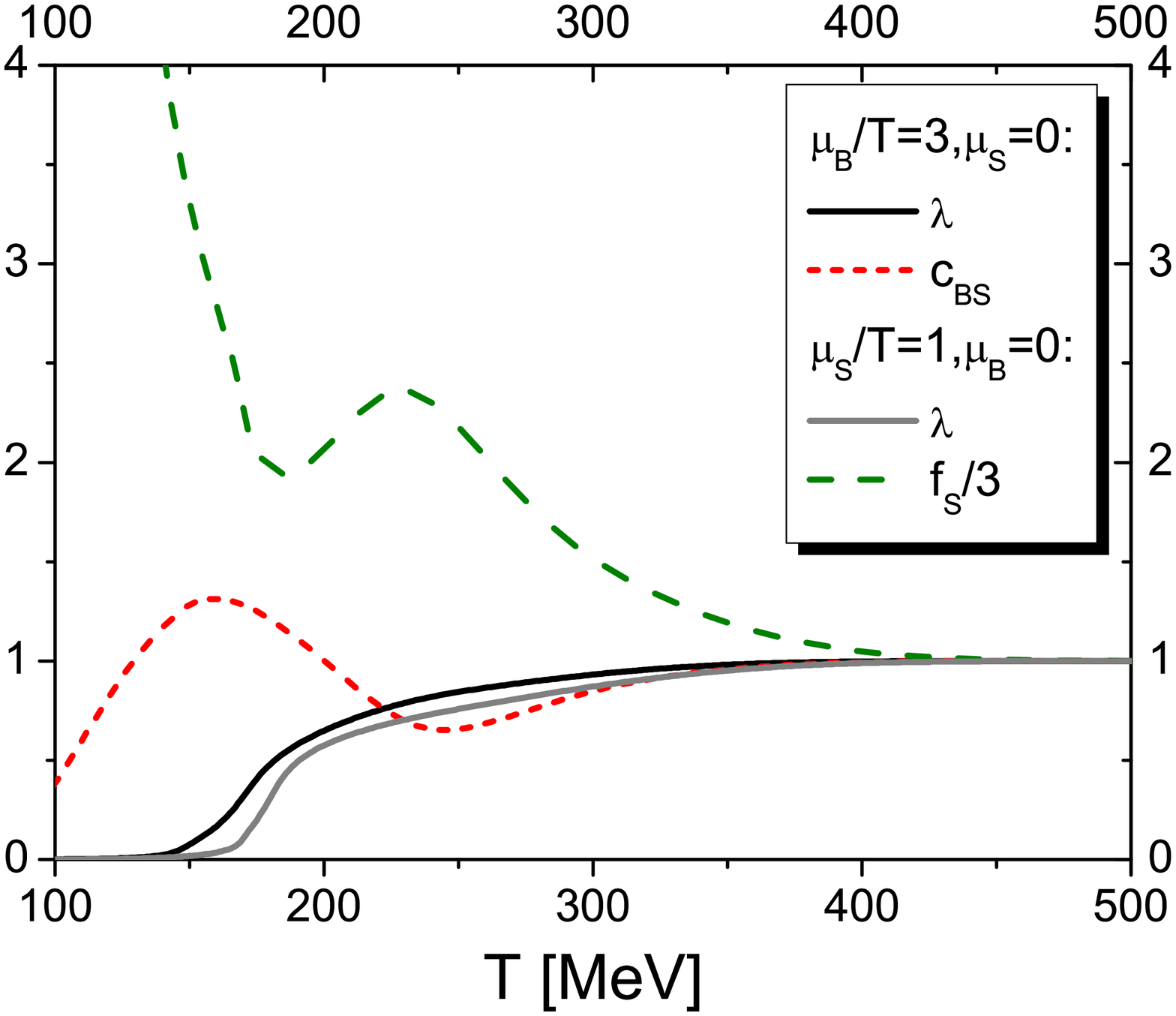}
 \caption{(Color online) Shown are the strangeness to baryon correlation coefficient $c_{BS}$ (red short dashed line) compared to the quark-gluon fraction $\lambda=e_{Quarks+Gluons}/e_{Tot}$ (black solid line) as a function of temperature for $\mu_B/T=3$ and $\mu_S=0$. The plot also shows the strangeness per baryon fraction $f_s$ (green dashed line) and the quark-gluon fraction $\lambda$ (grey solid line) as a function of temperature for $\mu_S/T=1$ and $\mu_B=0$.}
 \label{13}
\end{figure}

Figure \ref{11} presents our results on the order parameter of the chiral phase transition as a function of $\mu_B$ and $\mu_S$ at fixed temperature. The red lines indicate paths of constant values for $f_s=\rho_s/\rho_B$, the strangeness per baryon fraction. At the temperature $T=56$ MeV, the critical endpoint of the chiral phase transition was located at $\mu_B^{cep}\approx 1150$ MeV. We can observe that for increasing $f_s$, the change in the order parameter becomes steeper and the value of $T_{CEP}$ increases slightly to $T_{CEP}= 68$ MeV for $f_s=0.5$.
For a gas of deconfined quarks there is a strong correlation between the baryon number and strangeness. In a hadronic medium such a correlation is usually not trivial as strangeness can be found in mesons and baryons. These considerations led to the idea that the so called strangeness-baryon correlation factor $c_{BS}=-3\frac{\left\langle N_B N_S\right\rangle- \left\langle N_B \right\rangle \left\langle N_S\right\rangle}{\left\langle N_S^2 \right\rangle - \left\langle N_S \right\rangle^2}$ is sensitive to the deconfinement and/or chiral phase transition \cite{Koch:2005vg}. On the other hand the strangeness to baryon ratio $f_s$ should also be sensitive on any phase transition at finite baryon densities. On the lattice such quantities are usually calculated as functions of the expansion coefficients.
The information that can be extracted from these quantities is exemplified in figure \ref{13}. Here we show $c_{BS}$ as a function of temperature for $\mu_B/T=3$ and $\mu_S=0$. One can observe a distinct peak at $T\approx 150$ MeV $\Rightarrow \mu_B=450$ MeV. One can identify this peak with the crossover transition of the chiral condensate. Such a behavior of $c_{BS}$ has been predicted and also has been shown to exist in lattice data \cite{Schmidt:2009qq}. At higher temperatures the strangeness to baryon correlation approaches unity which resembles closely the behavior of the quark and gluon fraction $\lambda=e_{Quarks+Gluons}/e_{Tot}$ of the system. In comparison figure \ref{13} also shows the temperature dependence of $f_s$ at $\mu_S/T=1$ and $\mu_B=0$. This quantity is even more sensitive in the quark-gluon fraction as $c_{BS}$, while it seems to be not very sensitive to the chiral phase transition.

\section{Summary}
We presented results on the production of hypernuclear systems in high energy collisions of heavy ions. In particular we have investigated the production of hyperons in peripheral relativistic heavy ion collisions and their capture 
by the attractive potential of spectator residues. The absorption rate of hyperons in the excited spectators is shown to be quite substantial. This opens the possibility to study the phase transition in nuclear matter with a strangeness admixture and reveal information about the properties of hypernuclei, their binding energies, and, finally, $YN$ and $YY$ interactions.
In the second part of this work we discuss properties of the phase diagram at finite net-strange density within a SU(3) parity doublet model. We find that the location of the critical endpoint shifts to a slightly higher temperature for a finite net strangeness (lattice results). In particular the strangeness baryon correlation factor $c_{BS}$ and the strangeness per baryon fraction $f_s$ both show to be sensitive to the deconfined fraction on the system while $c_{BS}$ also shows a distinct peak at the chiral crossover at finite chemical potential.  

This work was supported by the Hessian LOEWE initiative Helmholtz International Center for FAIR, EMMI and used computational resources provided by the (L)CSC at Frankfurt.

\end{document}